\documentstyle[preprint,aps,epsf]{revtex}
\tightenlines
\begin{document}
\draft
\preprint{
}
\title{
Reanalysis of the Four-Quark Operators Relevant to $\Lambda_b$ Lifetime 
from QCD Sum Rule
}
\author{
Chao-Shang Huang$^{\:a}$, Chun Liu$^{\:b}$ and Shi-Lin Zhu$^{\:a}$
}
\vspace{0.5cm}
\address{
$^a$Institute of Theoretical Physics, Academia Sinica, P.O. Box 2735,\\ 
Beijing 100080, China\\
$^b$Korea Institute for Advanced Study, 207-43 Cheongryangri-dong,
Dongdaemun-gu,\\ 
Seoul 130-012, Korea
}
\maketitle
\thispagestyle{empty}
\setcounter{page}{1}
\begin{abstract}
$\Lambda_b$ matrix element of four-quark operator relevant to its 
lifetime is reanalyzed by QCD sum rule. The new ingredients introduced
are that (1) more vacuum condensates are considered; (2) different 
quark-hadron duality is adopted; and (3) the possible deviation
from the vacuum saturation assumption for the four quark condensates
is considered. With $\kappa_1=4$ the related hadronic parameter 
$r$ is calculated to be $(3.6\pm 0.9)$, and the lifetime ratio 
$\tau(\Lambda_b)/\tau(B^0)=(0.83\pm 0.04)$, 
which is consistent with the experimental data.  \\

   
\end{abstract}

\newpage

\section{Introduction}
\label{sec:introduction}

Heavy baryon lifetimes provide testing ground for the standard model, 
especially for QCD in some aspects, because they can be calculated 
systematically by heavy quark expansion \cite{1}.  Theoretically, if we 
do not assume the failure of the local duality assumption, the heavy 
hadron lifetime differences appear, at most, at the order of $1/m_Q^2$ 
\cite{2}.  Recent experimental result on the lifetime ratio of the 
$\Lambda_b$ baryon and $B$ meson is \cite{3} 
\begin{equation}
\label{1}
\frac{\tau(\Lambda_b)}{\tau(B^0)} = 0.79 \pm 0.06 \; ,
\end{equation}
which has shown some deviation from the theoretical expectation.
This is one of issues concerned with the tests of the heavy quark 
expansions and 
has drawn a lot of theoretical attention \cite{4,5,6,7,8}, and may 
imply the potential importance of the $O(1/m_b^3)$ effect for the above 
heavy baryon and heavy meson lifetime difference \cite{4,9,10}.

It is at the order $1/m_b^3$ that appear four-quark operators, 
whose contributions to decay widths are enhanced by a phase-space
factor of ${\cal O} (16\pi^2)$ with respect to the leading terms
in the operator product expansion. Consequently one should include
them in the prediction for non-leptonic decay rates \cite{155}, 
To the order of $1/m_b^3$, 
the lifetime ratio was calculated as follows \cite{4},
\begin{eqnarray}
\label{2}
\frac{\tau(\Lambda_b)}{\tau(B^0)} & = & 1 + 
\frac{\mu_{\pi}^2(\Lambda_b)-\mu_{\pi}^2(B)}{2m_b^2} - 
c_G\frac{\mu_G^2(B)}{m_b^2} \nonumber\\[3mm]
&& +\xi\{p_1B_1(m_b)+p_2B_2(m_b)+p_3\epsilon_1(m_b)+p_4\epsilon_2(m_b)
+[p_5+p_6\tilde{B}(m_b)]r(m_b)\} \; ,
\end{eqnarray}
where the term proportional to $\xi\equiv (f_B/200 {\rm MeV})^2$ 
arises from the contributions of four-quark operators.
The numerical values of the coefficients $c_G$
and $p_i$'s ($i=1-6$) have been calculated in Ref. \cite{4}.  
$\mu_{\pi}^2(H_b)$ and $\mu_G^2(B)$ are the averages of $b$-quark kinetic 
and chromomagnetic energy, respectively.  
To the order of $1/m_b^2$, the lifetime ratio is $0.98$.  At the scale
$m_b$, the values of $p_i$'s are $p_1=-0.003$, $p_2=0.004$, $p_3=-0.173$,
$p_4=-0.195$, $p_5=-0.012$, $p_6=-0.021$.  $B_1$, $B_2$, $\epsilon_1$, 
$\epsilon_2$, $r$ and $\tilde{B}$ are the parameterization of the 
hadronic matrix elements of the following four-quark operators,
\begin{eqnarray}
\label{3}
\langle\bar{B}|\bar{b}\gamma_{\mu}(1-\gamma_5)q\bar{q}\gamma^{\mu}
(1-\gamma_5)b|\bar{B}\rangle & \equiv & B_1f_B^2m_B^2 \; , \nonumber\\
\langle\bar{B}|\bar{b}(1-\gamma_5)q\bar{q}(1+\gamma_5)b
|\bar{B}\rangle & \equiv & B_2f_B^2m_B^2 \; , \nonumber\\
\langle\bar{B}|\bar{b}\gamma_{\mu}(1-\gamma_5)t_aq\bar{q}\gamma^{\mu}
(1-\gamma_5)t_ab|\bar{B}\rangle & \equiv & \epsilon_1f_B^2m_B^2 \; , 
\nonumber\\
\langle\bar{B}|\bar{b}(1-\gamma_5)t_aq\bar{q}(1+\gamma_5)t_ab
|\bar{B}\rangle & \equiv & \epsilon_2f_B^2m_B^2 \; ,
\end{eqnarray}
and 
\begin{eqnarray}
\label{4}
\displaystyle\frac{1}{2m_{\Lambda_b}}\langle\Lambda_b|
\bar{b}\gamma_{\mu}(1-\gamma_5)q\bar{q}\gamma^{\mu}(1-\gamma_5)b
|\Lambda_b\rangle & \equiv & \displaystyle-\frac{f_B^2m_B}{12}r \; , 
\nonumber\\[3mm]
\displaystyle\frac{1}{2m_{\Lambda_b}}\langle\Lambda_b|
\bar{b}(1-\gamma_5)q\bar{q}(1+\gamma_5)b|\Lambda_b\rangle & \equiv & 
\displaystyle-\tilde{B}\frac{f_B^2m_B}{24}r \; .
\end{eqnarray}
In Eq. (\ref{2}), the energy scale for the parameters is at $m_b$.  In Eq. 
(\ref{3}), generally the renormalization scale is arbitrary, and the 
parameters depend on it. It can be taken naturally at the low hadronic 
scale to apply the heavy quark expansion.

These parameters should be calculated by some nonperturbative QCD method.  
The QCD sum rule \cite{11}, which is regarded as a nonperturbative method 
rooted in QCD itself, has been used successfully to calculate the 
properties of various hadrons.  In Refs. \cite{5} and \cite{6}, the 
mesonic parameters $B_i$ and $\epsilon_i$ have been calculated by the QCD 
sum rules within the framework of heavy quark effective theory (HQET) 
\cite{12}.  The baryonic parameters $r$ and $\tilde{B}$ have been 
calculated in Ref. \cite{7}.  As a result, to the order of $1/m_b^3$, the 
theoretical calculation still cannot explain the experiment result 
(\ref{1}).

\section{Theoretical Consideration}
\label{sec:theore}

In this paper, we try to match the experimental data by considering more 
subtleties in the theoretical analysis.  
As analyzed in Ref. \cite{122}, the lifetime ratio (\ref{1}) depends 
crucially on the value of the baryonic parameter $r$.
Therefore, we shall carry out a reanalysis for baryonic 
parameter $r$ by the QCD sum rule.  We note that in the 
analysis of Ref. \cite{7}, more condensates could be included in.  They 
are gluon condensate and six-quark condensate.  It may be important because 
the dimension of the gluon condensate is four.  In spite of being dimension
nine, the six-quark condensate may also have significant contribution, 
because the diagram is not suppressed by any loop factor.  Experience from
the heavy baryon masses tells us that the pure tree diagram is important 
\cite{13,14,15,16}.  Therefore in our calculation, these two kinds of 
condensates will be considered, in addition to those considered in Ref. 
\cite{7}.  

The conclusion of $\tilde{B}=1$ does not change.  It follows from the 
valence quark approximation which is used both in Ref. \cite{7} and in our 
QCD sum rule analysis.

To calculate $r$, the following three-point Green's function is constructed,
\begin{equation}
\label{5}
\Pi(\omega, \omega') = i^2\int dxdye^{ik'\cdot x-ik\cdot y}\langle 0|
{\cal T}\tilde{j}^v(x) {\tilde O}(0) \bar{\tilde{j}^v}(y)|0\rangle \; ,
\end{equation}
where $\omega=v\cdot k$ and $\omega'=v\cdot k'$.  The $\Lambda_Q$ baryonic
current $\tilde{j}^v$ was given in Refs. \cite{13,14,15,16}, 
\begin{equation}
\label{6}
\tilde{j}^v = \epsilon^{abc}q_1^{Ta}C\gamma_5(a+b\not v)\tau
q_2^b h_v^c \; ,
\end{equation}
where $a$ and $b$ are certain constants which will be 
discussed later, $h_v$ is the heavy quark field in the HQET with velocity 
$v$, $C$ is the charge conjugate matrix, $\tau$ is the flavor matrix for 
$\Lambda_Q$, 
\begin{equation}
\label{7}
\tau = \frac{1}{\sqrt{2}}\left(\begin{array}{cc}
0  & 1 \\
-1 & 0
\end{array}
\right) \; .
\end{equation}
In Eq. (\ref{5}), ${\tilde O}$ denotes the four-quark operator
\begin{equation}
\label{8}
{\tilde O} = \bar{h_v}\gamma_{\mu}{1-\gamma_5\over 2}h_v
\bar{q}\gamma^{\mu}{1-\gamma_5\over 2}q \; .
\end{equation}
Note $<\Lambda_b |{\tilde O}|\Lambda_b > =-<\Lambda_b |O|\Lambda_b >$ 
in the valence quark approximation \cite{7}, where 
\begin{equation}
\label{8-1}
O= \bar{h_v}\gamma_{\mu}{1-\gamma_5\over 2}q
\bar{q}\gamma^{\mu}{1-\gamma_5\over 2} h_v \; .
\end{equation}
In terms of the hadronic expression, the parameter $r$ appears in the 
ground state contribution of $\Pi(\omega, \omega')$, 
\begin{equation}
\label{9}
\Pi(\omega, \omega') = {1\over 2}\frac{f_{\Lambda}^2\langle\Lambda_Q|O
|\Lambda_Q\rangle}{(\bar{\Lambda}-\omega)(\bar{\Lambda}-\omega')}
\frac{1+\not v}{2} + {\rm higher~~states} \; ,
\end{equation}
where the "higher states" denotes the contribution of resonances and 
continuum.  
$\langle\Lambda_Q|O|\Lambda_Q\rangle$ has appeared in Eq. (\ref{4}), 
$\bar{\Lambda}=m_{\Lambda_Q}-m_Q$ and the quantity $f_{\Lambda}$ is defined 
as 
\begin{equation}
\label{10}
\langle 0|\tilde{j}^v|\Lambda_Q\rangle \equiv f_{\Lambda} u \; ,
\end{equation}
with $u$ being the unit spinor in the HQET.  The QCD sum rule calculations
for $f_{\Lambda}$ were given in Refs. \cite{13,14,15,16}.  On the other hand,
this Green's function $\Pi(\omega, \omega')$ can be calculated in terms of 
quark and gluon language with vacuum condensates.  This establishes the sum 
rule.  For the resonance part of Eq. (\ref{9}), we adopt the assumption of
quark-hadron duality.

The calculation of $\Pi(\omega, \omega')$ in HQET is straightforward.  The 
fixed point gauge \cite{17} is used.  The tadpole diagrams in which the light
quark lines from the four-quark vertex are contracted have been subtracted.  
The following values of the condensates are used,
\begin{eqnarray}
\label{11}
\langle\bar{q}q\rangle & \simeq & -(0.23~ {\rm GeV})^3 \; , \nonumber\\
\langle\alpha_sGG\rangle & \simeq & (0.075\pm 0.015)~ {\rm GeV}^4 \; , \nonumber\\
\langle g\bar{q}\sigma_{\mu\nu}G^{\mu\nu}q\rangle & \equiv &
m_0^2\langle\bar{q}q\rangle \; ,
~~~m_0^2\simeq0.8~{\rm GeV}^2 \; .
\end{eqnarray}
Note we have adopted the new value for the gluon condensate 
from the recent analysis of the heavy quarkonium spectrum \cite{20}.
However, our final results are not very sensitive to the gluon 
condensate since its contribution is only about $6\%$ of the whole sum 
rule as shall see later.

Except for the quark gluon mixed condensate, 
our calculation would be consistent with that of Ref. \cite{7} if the 
gluon and six-quark condensates were omitted.  
While the calculation can be justified
if ($-\omega$) and ($-\omega'$) are large, however the hadron ground state 
property should be obtained at small ($-\omega$) and ($-\omega'$).  These 
contradictory requirements are achieved by introducing double Borel 
transformation for $\omega$ and $\omega'$.  It is defined as 
\begin{equation}
\label{12}
\hat{B} = \lim_{\begin{array}{c}
-\omega\to\infty\\n\to\infty\\
\tilde{\tau}\equiv\frac{-\omega}{n}~ {\rm fixed} 
\end{array}}
\lim_{\begin{array}{c}-\omega'\to\infty\\m\to\infty\\ 
\tilde{\tau}'\equiv\frac{-\omega'}{m}~ {\rm fixed}
\end{array}}
\frac{(-\omega )^{n+1}}{n!}\left(\frac{d}{d\omega }\right)^n
\frac{(-\omega')^{m+1}}{m!}\left(\frac{d}{d\omega'}\right)^m \; .
\end{equation}
There are two Borel parameters $\tilde{\tau}$ and $\tilde{\tau}'$.  They 
appear symmetrically, so $\tilde{\tau}=\tilde{\tau}'=2T$ is taken in the 
following analysis. The reason for the factor $2$ is similar to that 
explained in the Ref. \cite{123}.

\section{Duality Assumption}
\label{sec:duality}

Generally the duality is to simulate the higher states by the 
whole quark and gluon contribution above some threshold energy 
$\omega_c$.  The whole  
contribution of the three-point correlator $\Pi(\omega, \omega')$ 
can be expressed by the dispersion relation,
\begin{equation}
\label{13}
\Pi(\omega, \omega') = \frac{1}{\pi}\int_0^{\infty}d\nu\int_0^{\infty}
d\nu'\frac{{\rm Im}\Pi(\nu, \nu')}{(\nu-\omega)(\nu'-\omega')} \; .
\end{equation}
With the redefinition of the integral variables
\begin{eqnarray}
\label{14}
\nu_+ & = & \displaystyle\frac{\nu+\nu'}{2} \; , \nonumber\\
\nu_- & = & \displaystyle\frac{\nu-\nu'}{2} \; ,
\end{eqnarray}
the integration becomes
\begin{equation}
\label{15}
\int_0^{\infty}d\nu\int_0^{\infty}d\nu'... = 
2\int_0^{\infty}d\nu_+\int_{-\nu_+}^{\nu_+}d\nu_- ... \; .
\end{equation}
It is in $\nu_+$ that the quark-hadron duality is assumed \cite{12},
\begin{equation}
\label{16}
{\rm higher~~states} = \frac{2}{\pi}\int_{\omega_c}^{\infty}d\nu_+
\int_{-\nu_+}^{\nu_+}d\nu_-
\frac{{\rm Im}\Pi(\nu, \nu')}{(\nu-\omega)(\nu'-\omega')} \; .
\end{equation}
This kind of assumption was suggested in calculating the Isgur-Wise 
function in Ref. \cite{123} and was argued for in Ref. \cite{22a}.
As pointed out in \cite{22a,12}, in calculating three-point functions 
the duality is valid after integrating the spectral density over the 
"off-diagonal" variable $\nu_-=\frac{1}{2}(\nu-\nu')$.

The sum rule for $\langle\Lambda_Q|{\tilde O}|\Lambda_Q\rangle$ 
after the integration with the variable $\nu_-$ is obtained as
\begin{eqnarray}
\label{17}
{(a+b)^2\over 2}
f_{\Lambda}^2\exp\left(-\frac{\bar{\Lambda}}{T}\right)
\langle\Lambda_Q|{\tilde O}|\Lambda_Q\rangle & = & 
\displaystyle \int_0^{\omega_c}d\nu \exp\left(-\frac{\nu}{T}\right)
\{ \frac{a^2+b^2}{840\pi^6} \nu^8 
-\frac{ab}{6\pi^4}\nu^5 \langle\bar{q}q\rangle \nonumber\\[3mm]
&& \displaystyle +\frac{3(a^2+b^2)}{2048\pi^6}\nu^4 \langle g^2_sG^2 \rangle
+ \frac{5ab}{48\pi^4} m_0^2\langle\bar{q}q\rangle \nu^3 \nonumber\\[3mm]
&&\displaystyle +\kappa_1 \frac{17(a^2+b^2)}{96\pi^2}
\langle\bar{q}q\rangle^2 \nu^2 \}
-\kappa_2 \frac{ab}{144} \langle\bar{q}q\rangle^3 \;,
\end{eqnarray}
where $\kappa_1$, $\kappa_2$ are the scale parameters used to indicate 
the possible deviation from the factorization assumption for the 
four- and six-quark condensates. $\kappa_{1,2}=1$ corresponds to
the vacuum saturation approximation. $\kappa_1 =(3\sim 8)$ is 
often introduced in order to include the nonfactorizable 
contribution and to fit the data \cite{21}. There is no discussion
of $\kappa_2$ in literature so we use $\kappa_2 =1$.

As argued in \cite{15}, the choice $a=b=1$ tends to yield the 
optimal interpolating current for the $\Lambda_b$ baryon in HQET.
We shall adopt $a=b=1$ in our numerical analysis.

The parameters $f_{\Lambda}, {\bar\Lambda}$ etc 
were obtained by the HQET sum rule analysis of 
two-point correlator \cite{13,14,15,16},
\begin{equation}
\label{19}
{(a+b)^2\over 2}f_{\Lambda}^2e^{-\bar{\Lambda}/T} 
= \int_0^{\omega_c}d\nu e^{-\nu/T} [\frac{a^2 +b^2}{20\pi^4}\nu^5
-\frac{2ab\langle\bar{q}q\rangle}{\pi^2}(\nu^2 -{m_0^2\over 16})]
\; .\\[3mm]
\end{equation}
We have not included $\alpha_s$ corrections in Eq. (\ref{19}), 
because they are also neglected in the sum rule for $r$ 
(\ref{17}).
The values of the parameters are
$f_{\Lambda}^2 =(2.9\pm 0.5)\times 10^{-2}$ GeV$^3$, 
$\bar\Lambda =(0.9\pm 0.1) $GeV with 
the threshold $\omega_c$ to be $(1.2\pm 0.2) $GeV and 
the Borel parameter $T$ in the window ($0.2-0.4$) GeV.

In order to minimize the dependence of the parameters we 
divide Eq. (\ref{17}) by Eq. (\ref{19}) to extract
$\langle\Lambda_Q|{\tilde O}|\Lambda_Q\rangle$.
The variation of the matrix element with $\omega_c$ 
for two values of $\kappa_1=1, 4$ is given in Fig. 1.  
The value of $\omega_c$ is $(1.2\pm 0.1)$ GeV.  
The sum rule window is $T=(0.15 - 0.35 )$ GeV, which is almost 
the same as that in the two-point correlator sum rule.
We obtain for $\kappa_1 =4$
\begin{equation}
\label{20}
\langle\Lambda_Q|{\tilde O}|\Lambda_Q\rangle = (1.6\pm 0.4)\times 
10^{-2} \mbox{GeV}^3 \; .
\end{equation}
By taking $f_B=200$ MeV,
\begin{equation}
\label{21}
r = (3.6\pm 0.9) \; .
\end{equation}
If we use $\kappa_1=1$, we get 
\begin{equation}
\label{20-a}
\langle\Lambda_Q|{\tilde O}|\Lambda_Q\rangle = (5.5\pm 1.0)\times 
10^{-3} \mbox{GeV}^3 \; ,
\end{equation}
\begin{equation}
\label{21-a}
r = (1.3\pm 0.3) \; .
\end{equation}

In our numerical calculation, the contribution of the perturbative term is
smaller than those of the condensates due to small $T$ in the duality window
and high dimensionality of the spectral function ( $\nu^8$ in eq. (18) ).
There are several known examples where the perturbative part has small 
contribution in the sum rule \cite{14,15,qd}, 
without affecting the reliability of the results.  
Nevertheless, a hierarchical structure among various condensate terms 
exists in those cases and the present case. In the present case, because of
the cancellation between the contributions of the quark condensate and the 
mixed 
condensate, as well as the smallness of the contributions of the gluon and 
the six-quark condensates,
the influence of the four-quark condensate is significant and is enhanced
by assuming the deviation from the vacuum saturation.

In Ref. \cite{7} the authors used the following duality assumption,
\begin{equation}
\label{16-1}
{\rm higher~~states} = \frac{1}{\pi}\int^{\infty}_{\omega_c} d\nu
\int^{\infty}_{\omega_c}d \nu' 
\frac{{\rm Im}\Pi(\nu, \nu')}{(\nu-\omega)(\nu'-\omega')} \; .
\end{equation}
In their calculation the matrix element 
$\langle\Lambda_Q|{\tilde O}|\Lambda_Q\rangle$ increases from 
$0.4 \times 10^{-3} $GeV$^3$ to $1.2 \times 10^{-3} $GeV$^3$
in the working region of the Borel parameter 
when $\omega_c$ varies from $1.1$ GeV to $1.3$ GeV. 
In other words, their analysis is very sensitive to the 
continuum threshold, which might imply the above duality assumption
is not good. In contrast, the dependence on $\omega_c$ is not 
so strong in our approach.

The $1/m_b$ corrections to the above results can be analyzed in principle.
While having little influence on our above calculation, they formally 
belong to the $O(1/m_b^4)$ effects to the $\Lambda_b$ lifetime.  It should
be noted that the value of $r$ we have obtained above is at some hadronic 
scale, other than the scale $m_b$, because we have been working in the framework
of the HQET, in which the natural scale is $\mu_{had}\ll m_b$.  The 
renormalization group evolution of the relevant operators was calculated in 
Ref. \cite{19,4}.  Information on parameters $\tilde{B}$ and $r$ at scale 
$m_b$ is necessary to obtain the lifetime ratio of Eq. (\ref{2}).  By choosing
$\alpha_s(\mu_{had})=0.5$ (corresponding to $\mu_{had}\sim0.67$ GeV), 
Ref. \cite{4} gives
\begin{eqnarray}
\label{22}
r(m_b) & \simeq & [1.54 + 0.18 \tilde{B}(\mu_{had})]r(\mu_{had}) \; , 
\nonumber\\[3mm]
\tilde{B}(m_b) & \simeq & 
\frac{\tilde{B}(\mu_{had})}{1.54 + 0.18\tilde{B}(\mu_{had})} \; .
\end{eqnarray}
That is 
\begin{equation}
\tilde{B}(m_b)  \simeq  0.58 \;,
\end{equation}
\begin{equation}
r(m_b) \simeq  (6.2\pm 1.6) 
\end{equation}
for $\kappa_1=4$ or
\begin{equation}
r(m_b) \simeq  (2.3\pm 0.6) 
\end{equation}
for $\kappa_1=1$ from Eqs. (\ref{21}) and (\ref{21-a}).

The $\Lambda_b$ and $B^0$ lifetime ratio given in Eq. (\ref{2}) is expressed
specifically as
\begin{eqnarray}
\label{24}
\frac{\tau(\Lambda_b)}{\tau(B^0)} & = & 0.98-0.17\epsilon_1(m_b)
+0.20\epsilon_2(m_b)-(0.013+0.022\tilde{B}(m_b))r(m_b) \nonumber\\
& = & (0.83\pm 0.04) 
\end{eqnarray}
for $\kappa_1=4$ or 
\begin{equation}
\label{24-a}
\frac{\tau(\Lambda_b)}{\tau(B^0)} =(0.93\pm 0.02) 
\end{equation}
for $\kappa_1=1$. 
Where the values $\epsilon_1(m_b)=-0.08$ and $\epsilon_2(m_b)=-0.01$ have 
been taken from 
the QCD sum rules \cite{5}.  From Eq. (\ref{24-a}), we see that with the 
vacuum saturation ($\kappa_1=1$), although $r$ is enhanced by about six 
times compared to that in Ref. \cite{7},
it is still not large enough to account for the data Eq. (1).  The
life time ratio between $\Lambda_b $ and $B$ mesons can be explained
if we also take into account the nonfactorizable contribution of the 
four-quark condensate, as can be seen from Eq. (\ref{24}).

\section{Summary and Discussion}
\label{sec:summary}

In summary, we have reanalyzed the QCD sum rule for the $\Lambda_b$ matrix
element of the four-quark operator relevant to the lifetime of $\Lambda_b$.  
Compared to the previous analysis \cite{7}, the new ingredients we have 
introduced are that (i) more condensates are considered; (ii) different 
quark-hadron duality is adopted and (iii) the possible deviation
from the vacuum saturation assumption for the four quark condensates
is considered. Of these ingredients, the last two points have more
important influence on numerical results than the first point by noting the 
effect due to more condensates is small ($\sim 10\%$).
The second point is more essential than the third point.  Note that due to 
the second point, $r$ is about six times enhanced by comparing the result of
Eq. (\ref{21-a}) with that in Ref. \cite{7}.  And due to the third point, 
$r$ is about three times enhanced by comparing the results of Eqs. 
(\ref{21-a}) and (\ref{21}).  With fixed duality assumption for the sum 
rule, our result shows that the baryonic parameter $r$ is 
significantly dependent on the nonfactorizable effect of 
the four-quark condensate and $r(m_b) =(2.3\pm 0.6)$ 
for $\kappa_1=1$ and $r(m_b) =(6.2\pm 1.6)$ for $\kappa_1=4$.
With this latter value the $\Lambda_b$ baryon and $B^0$ meson 
lifetime ratio has been 
calculated to be $\tau(\Lambda_b)/\tau(B^0)=(0.83\pm 0.04)$, 
which is close to the experimental data 
$\tau(\Lambda_b)/\tau(B^0)=(0.79\pm 0.06)$. 
From the above results we draw the conclusion that under the 
local duality assumption it is possible to resolve the issue 
of the $\Lambda_b$ baryon and $B^0$ meson lifetime ratio 
in the framework of heavy quark expansion.

\acknowledgments

C.-S.H. and S.-L.Z. were supported in part by the Natural Science
Foundation of China.

\newpage

{\Large \bf Figure captions}\\

Fig. 1.  Sum rule for $\langle\Lambda_Q|{\tilde O}|\Lambda_Q\rangle$, 
where ${\tilde O}$ has been given in Eq. (\ref{8}). 
From top to bottom these curves correspond to $\omega_c =1.3, 1.2, 
1.1$ GeV respectively. The sum rule window is $T=(0.15-0.35)$ GeV. 

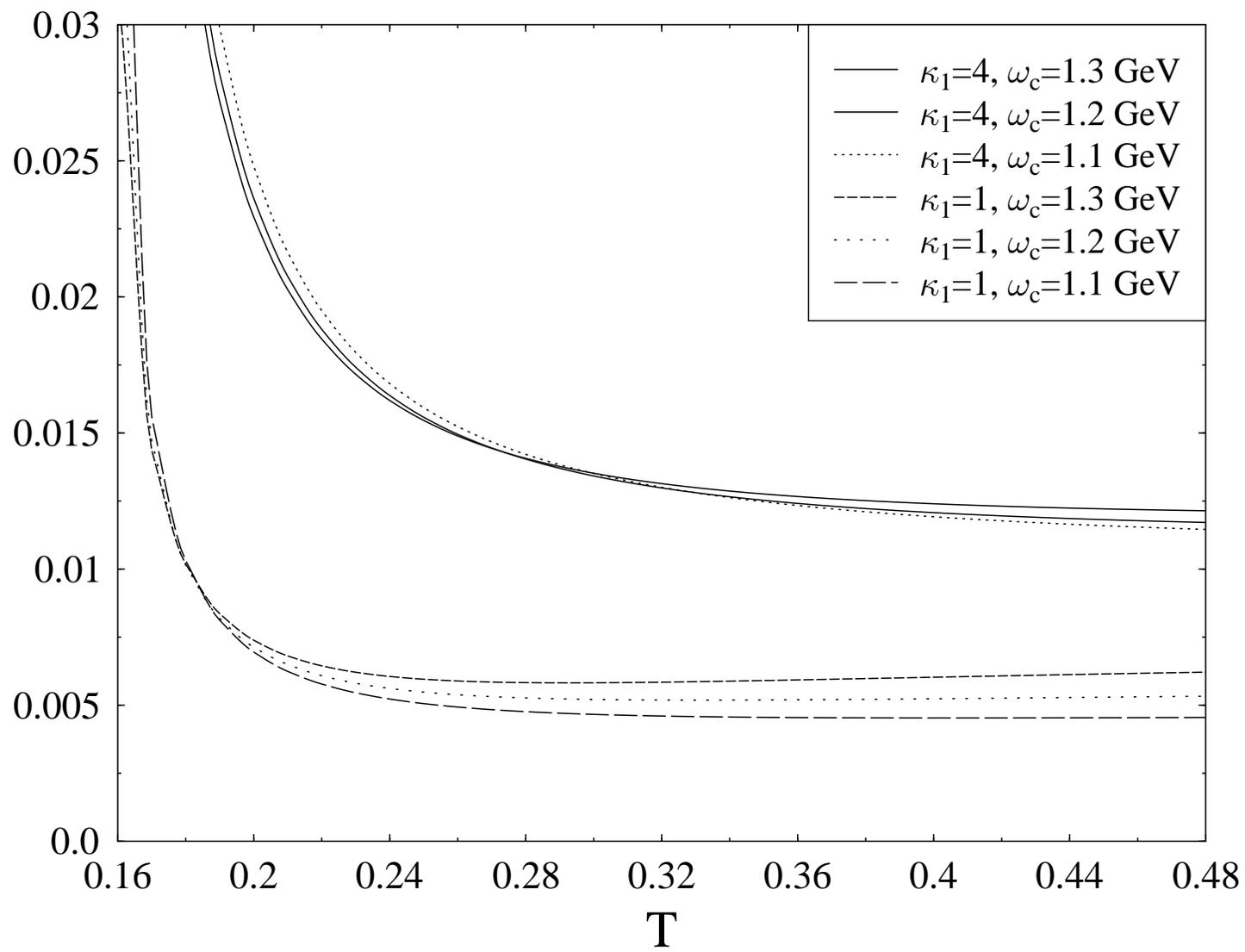
\end{document}